\documentstyle[12pt,procsla,epsfig]{article}
\newcommand{\nd}{\noindent}
\newcommand{\beq}{\begin{equation}}
\newcommand{\eeq}{\end{equation}}
\newcommand{\barr}{\begin{eqnarray}}
\newcommand{\earr}{\end{eqnarray}}
\newcommand{\ba}{\begin{array}}
\newcommand{\ea}{\end{array}}
\newcommand{\bfp}{\mbox{\boldmath $p$}}
\newcommand{\bfk}{\mbox{\boldmath $k$}}
\newcommand{\bfy}{\mbox{\boldmath $y$}}
\newcommand{\la}{\lambda}

\newcommand{\pup}{p^\uparrow}

  \catcode`\@=11
  \long\def\@makefntext#1{
  \protect\noindent \hbox to 3.2pt {\hskip-.9pt
  $^{{\ninerm\@thefnmark}}$\hfil}#1\hfill}              

  \def\@makefnmark{\hbox to 0pt{$^{\@thefnmark}$\hss}}  

  \def\ps@myheadings{\let\@mkboth\@gobbletwo
  \def\@oddhead{\hbox{}
  \rightmark\hfil\ninerm\thepage}
\def\@oddfoot{}\def\@evenhead{\ninerm\thepage\hfil
  \leftmark\hbox{}}\def\@evenfoot{}
  \def\sectionmark##1{}\def\subsectionmark##1{}}

\def\sumint{\nostrocostruttino \sum \over {\displaystyle\int}}
\def\lsim{\mathrel{\rlap{\lower4pt\hbox{\hskip1pt$\sim$}}\raise1pt\hbox{$<$}}} 
\def\gsim{\mathrel{\rlap{\lower4pt\hbox{\hskip1pt$\sim$}}\raise1pt\hbox{$>$}}} 
\def\nostrocostruttino#1\over#2{\mathrel{\mathop{\kern 0pt \rlap 
{\hbox{$#1$}}} \hbox{\kern-.135em $#2$}}}
\begin{document}
\begin{flushright}
DFTT 12/96 \\
INFNCA-TH9603 \\
hep-ph/9604397 \\
April 1996
\end{flushright}

  \vspace*{0.6cm}

\centerline{\normalsize\bf SINGLE SPIN ASYMMETRIES IN}
  \baselineskip=16pt
  \centerline{\normalsize\bf INCLUSIVE HADRON PRODUCTION
\footnote{$\,$ Talk delivered by M.~Anselmino at the Adriatico Research
Conference on ``Trends in Collider Spin Physics'', Trieste, Italy,
December 1995, and at the International Conference on Hadron
Structure: High-Energy Interactions: Theory and Experiment
(Hadron Structure 96), Stara Lesna, Slovak Republik, February 1996.}}

  \vspace*{0.6cm}
  \centerline{\footnotesize MAURO ANSELMINO, MARIA ELENA BOGLIONE}
  \baselineskip=13pt
  \centerline{\footnotesize\it 
 Dipartimento di Fisica Teorica, Universit\`a di Torino}
  \baselineskip=12pt
  \centerline{\footnotesize\it and INFN, Sezione di Torino}
  \baselineskip=12pt
  \centerline{\footnotesize\it Via P. Giuria 1, I-10125 Torino, Italy}
  \centerline{\footnotesize E-mail: anselmino@torino.infn.it,
   boglione@torino.infn.it}
  \vspace*{0.3cm}
  \centerline{\footnotesize and}
  \vspace*{0.3cm}
  \centerline{\footnotesize FRANCESCO MURGIA}
  \baselineskip=13pt
  \centerline{\footnotesize\it INFN, Sezione di Cagliari}
  \baselineskip=12pt
  \centerline{\footnotesize\it Via Ada Negri 18, I-09127 Cagliari, Italy }
  \centerline{\footnotesize E-mail: murgia@vaxca.ca.infn.it}

  \vspace*{0.9cm}
  \abstracts{The general formalism to describe single spin asymmetries
in hadron-hadron high energy and large $p_T$ inclusive production
within the QCD-improved parton model and assuming the factorization
theorem to hold for higher twist contributions is discussed.
Non zero values of the single spin asymmetries originate from and
reveal non perturbative universal properties of quarks: the quark
distribution and fragmentation analysing powers, which need not be
zero, provided the quark intrinsic motion is taken into account.
A simple model is constructed which reproduces the main features of
the data on the single spin asymmetries observed in inclusive pion 
production in $pp$ collisions.}

  \normalsize\baselineskip=15pt
  \setcounter{footnote}{0}
  \renewcommand{\thefootnote}{\alph{footnote}}

\vskip1.0truecm

According to the QCD factorization theorem the differential cross-section 
for the hard scattering of a polarized hadron $A$ with spin $S_A$ off 
another polarized hadron $B$ with spin $S_B$, resulting in the inclusive 
production of a hadron $C$ with energy $E_C$ and three-momentum 
$\bfp_C$, $A,S_A + B,S_B \to C + X$, can be written as \cite{col1,col2,col3}
\barr
\frac{E_C\, d\sigma^{A,S_A+B,S_B \to C+X}} {d^{3} \bfp_C} &=&
\sum_{a,b,c,d} \>
\sum_{\la^{\,}_a, \la^{\prime}_a; \la^{\,}_b, \la^{\prime}_b;
\la^{\,}_c, \la^{\prime}_c; \la^{\,}_d; \la^{\,}_C}
\int {dx_a \, dx_b \over \pi z} \frac{1}{16\pi\hat s^2}\label{gennok} \\  
& &\mbox{\hskip-3.0truecm}\times\,
\rho_{\la^{\,}_a, \la^{\prime}_a}^{a/A,S_A} \,
f_{a/A}(x_a) \> \rho_{\la^{\,}_b, \la^{\prime}_b}^{b/B,S_B} \,
\,f_{b/B}(x_b) \,
\hat M_{\la^{\,}_c, \la^{\,}_d; \la^{\,}_a, \la^{\,}_b} \, 
\hat M^*_{\la^{\prime}_c, \la^{\,}_d; \la^{\prime}_a, \la^{\prime}_b} \,
D_{\la^{\,}_C,\la^{\,}_C}^{\la^{\,}_c,\la^{\prime}_c}(z) \,,
\nonumber
\earr
where $f_{i/I}(x_i)$ is the number density of partons $i$ with momentum 
fraction $x_i$ inside hadron $I$ ($i = a,b; \> I = A,B$) and 
$\rho_{\la^{\,}_i, \la^{\prime}_i}^{i/I,S_I}(x_i)$ 
is the helicity density matrix of parton $i$ inside the polarized hadron $I$.
The $\hat M_{\la^{\,}_c, \la^{\,}_d; \la^{\,}_a, \la^{\,}_b}$'s 
are the helicity amplitudes for the elementary process $ab \to cd$;
if one wishes to consider higher order (in $\alpha_s$) contributions also 
elementary processes involving more partons should be included.    
$D_{\la^{\,}_C,\la^{\prime}_C}^{\la^{\,}_c,\la^{\prime}_c}(z)$ 
is the product of {\it fragmentation amplitudes}
\beq
D_{\la^{\,}_C,\la^{\prime}_C}^{\la^{\,}_c,\la^{\prime}_c} 
= \> \sumint_{X, \la_{X}} {\cal D}_{\la^{\,}_{X},\la^{\,}_C;\la^{\,}_c} \,
{\cal D}^*_{\la^{\,}_{X},\la^{\prime}_C; \la^{\prime}_c}\, ,
\label{framp}
\eeq
where the $\sumint_{X, \la_{X}}$ stays for a spin sum and phase space
integration of the undetected particles, considered as a system $X$.
The usual unpolarized fragmentation function $D_{C/c}(z)$, {\it i.e.} 
the density number of hadrons $C$ resulting from the fragmentation of 
an unpolarized parton $c$ and carrying a fraction $z$ of its momentum
is given by
\beq
D_{C/c}(z) = {1\over 2} \sum_{\la^{\,}_c,\la^{\,}_C} 
D_{\la^{\,}_C,\la^{\,}_C}^{\la^{\,}_c,\la^{\,}_c}(z)\,.
\label{fr}
\eeq

For simplicity of notations we have not shown in Eq. (\ref{gennok})
the $Q^2$ scale dependences in $f$ and $D$; the variable $z$ is related to 
$x_a$ and $x_b$ by the usual imposition of energy momentum conservation
in the elementary 2 $\to$ 2 process \cite{bla}, $z = -(x_a t + x_b u)
/x_a x_b s$ where $s,t,u$ are the Mandelstam variables for the 
overall process $AB \to CX$ whereas $\hat s, \hat t, \hat u$ are those 
for the elementary process $ab \to cd$. The corresponding amplitudes 
are normalized so that 
\beq
\frac{d\hat\sigma}{d\hat t} = \frac{1}{16\pi\hat s^2}\frac{1}{4}
\sum_{\la^{\,}_a, \la^{\,}_b, \la^{\,}_c, \la^{\,}_d}
|\hat M_{\la^{\,}_c, \la^{\,}_d; \la^{\,}_a, \la^{\,}_b}|^2\,.
\label{norm}
\eeq

Eq. (\ref{gennok}) holds at leading twist and large $p_T$ values of the 
produced pion; the intrinsic $\bfk_\perp$ of the partons have 
been integrated over and collinear configurations dominate both the 
distribution functions and the fragmentation processes; one can then see
that, in this case, there cannot be any single spin asymmetry. 

Suppose, in fact, that hadron $B$ is not polarized so that 
Eq. (\ref{gennok}) reads
\barr
\frac{E_C\, d\sigma^{A,S_A+B \to C+X}} {d^{3} \bfp_C} &=& 
\sum_{a,b,c,d} \>
\sum_{\la^{\,}_a, \la^{\prime}_a; \la^{\,}_b; 
\la^{\,}_c, \la^{\prime}_c; \la^{\,}_d; \la^{\,}_C} {1 \over 2}
\int {dx_a \, dx_b \over \pi z} \frac{1}{16\pi\hat s^2}\label{bunp1} \\  
& &\mbox{\hskip-1.0truecm}\times \,
\rho_{\la^{\,}_a, \la^{\prime}_a}^{a/A,S_A} \,
f_{a/A}(x_a) \, f_{b/B}(x_b) \,
\hat M_{\la^{\,}_c, \la^{\,}_d; \la^{\,}_a, \la^{\,}_b} \, 
\hat M^*_{\la^{\prime}_c, \la^{\,}_d; \la^{\prime}_a, \la^{\,}_b} \,
D_{\la^{\,}_C,\la^{\,}_C}^{\la^{\,}_c,\la^{\prime}_c}(z) \, .
\nonumber
\earr

Then total angular momentum conservation in the (forward) fragmentation process 
[see Eq. (\ref{framp})] implies $\la^{\,}_c = \la^{\prime}_c$; this, in turns,
together with helicity conservation in the elementary processes, implies
$\la^{\,}_a = \la^{\prime}_a$. If we further notice that, by parity invariance,
$\sum_{\la^{\,}_C} D_{\la^{\,}_C,\la^{\,}_C}^{\la^{\,}_c,\la^{\,}_c} = 
D_{C/c}$ independently of $\la_c$ and that
\beq
\sum_{\la^{\,}_b, \la^{\,}_c, \la^{\,}_d} 
|\hat M_{\la^{\,}_c, \la^{\,}_d; \la^{\,}_a, \la^{\,}_b} |^2 =
2\left(16\pi\hat s^2\right)\frac{d\hat\sigma}{d\hat t}\, ,
\label{parity}
\eeq
we remain with 
$\sum_{\la^{\,}_a} \rho_{\la^{\,}_a, \la^{\,}_a}^{a/A,S_A} = 1$. Moreover,
in the absence of intrinsic $\bfk_\perp$ and initial state interactions,
the parton density numbers $f_{a/A}(x_a)$ cannot depend on the hadron 
spin and any spin dependence disappears from Eq. (\ref{gennok}), so that
\barr
\frac{E_C\, d\sigma^{A,S_A + B \to C+X}} {d^{3} \bfp_C} &=& 
\sum_{a,b,c,d} \>
\int {dx_a \, dx_b \over \pi z}\, f_{a/A}(x_a) \,f_{b/B}(x_b)
\nonumber \\
&\times& \frac{d\hat\sigma^{a+b\to c+d}}{d\hat t}\,
D_{C/c}(z) \; = \; \frac{E_C \, d\sigma^{unp}} {d^{3} \bfp_C}\, .
\label{bunp2}
\earr

This implies that the single spin asymmetry, 
\barr
A_N & \equiv & \frac{d\sigma^{A^\uparrow\,B\to C\,X}(\bfp_T) -
d\sigma^{A^\downarrow\,B\to C\,X}(\bfp_T)}
{d\sigma^{A^\uparrow\,B\to C\,X}(\bfp_T) +
d\sigma^{A^\downarrow\,B\to C\,X}(\bfp_T)} \nonumber \\
& = & \frac{d\sigma^{A^\uparrow\,B\to C\,X}(\bfp_T) -
d\sigma^{A^\uparrow\,B\to C\,X}(-\bfp_T)}
{d\sigma^{A^\uparrow\,B\to C\,X}(\bfp_T) +
d\sigma^{A^\uparrow\,B\to C\,X}(-\bfp_T)} \, ,
\label{asym}
\earr
is zero. $\uparrow (\downarrow)$ refers to hadron $A$ spin orientation
up (down) with respect to the production plane; any other spin 
orientation with no up or down component would result in a vanishing
asymmetry due to parity invariance.

Eq. (\ref{gennok}) can be generalized with the inclusion of intrinsic 
$\bfk_\perp$ \cite{col2} and this can avoid the above conclusion; for example 
\cite{col2}, the observation of a non zero $\bfk_\perp$ of a final particle 
$C$ with respect to the axis of the jet generated by parton $c$ does not imply  
any more $\la^{\,}_c = \la^{\prime}_c$ and allows a non zero value of the
asymmetry through the {\it quark fragmentation analysing power}
\barr
A_{C/q} & = & \frac{D_{C/q,s_q}(z,\bfk_\perp)-
D_{C/q,-s_q}(z,\bfk_\perp)}{D_{C/q,s_q}(z,\bfk_\perp)+
D_{C/q,-s_q}(z,\bfk_\perp)} \nonumber \\
& = & \frac{D_{C/q,s_q}(z,\bfk_\perp)-
D_{C/q,s_q}(z,-\bfk_\perp)}{D_{C/q,s_q}(z,\bfk_\perp)+
D_{C/q,s_q}(z,-\bfk_\perp)} \, .
\label{acq}
\earr

A non zero value of the above quantity -- the analysing power or single 
spin asymmetry for the quark fragmentation process -- is allowed by parity 
invariance for quark spin orientations perpendicular to the $q$-$C$ plane 
and is allowed by time-reversal invariance due to the soft interactions 
of the fragmenting quark with external fields.

This idea was exploited in Ref.~5 to explain the single spin  
asymmetries observed in $\pup p \to \pi X$ processes \cite{ada2}
and where, essentially, it is assumed that parton 
$c$ is produced in the forward direction and the final hadron $p_T$ is due 
to its transverse $k_\perp$ inside the jet. One cannot expect such a model 
to work at large $p_T$, as the results clearly show.

Another possible $\bfk_\perp$ effect, suggested by Sivers \cite{siv1,siv2}, 
may originate in the distribution functions. Similarly to the Collins effect
in the fragmentation process a dependence on the hadron spin may remain in
the $\bfk_\perp$ dependent quark distribution function
$\tilde f_{q/A^\uparrow}(x, \bfk_{\perp})$, so that the difference 
\barr
\Delta^{\!N}\tilde f_{q/A^\uparrow}(x,\bfk_{\perp})
&\equiv& \tilde f_{q/A^\uparrow}(x,\bfk_{\perp})
- \tilde f_{q/A^\downarrow} (x,\bfk_{\perp}) \nonumber \\
&=& \tilde f_{q/A^\uparrow}(x,\bfk_{\perp}) -
\tilde f_{q/A^\uparrow} (x,-\bfk_{\perp}) 
\label{isiv2}
\earr
can be different from zero. $\tilde f_{q/A^\uparrow}(x, \bfk_\perp)$
is the number density of partons $a$ with momentum fraction $x_a$ 
and intrinsic transverse momentum $\bfk_\perp$ in a transversely polarized
hadron $A$.

Taking intrinsic $\bfk_\perp$ into account Eq. (\ref{bunp2}) becomes
\barr
\frac{E_C\, d\sigma^{A^\uparrow B \to C X}} {d^{3} \bfp_C} &=& 
\sum_{a,b,c,d} \>
\int {d^2\bfk_\perp dx_a \, dx_b \over \pi z} \nonumber \\  
&\times& \!\!\!\tilde f_{a/A^\uparrow}(x_a,\bfk_\perp)\,f_{b/B}(x_b) \,
\frac{d\hat\sigma}{d\hat t}(\bfk_\perp)\,
D_{C/c}(z)\, ,
\label{dskp}
\earr
so that, adopting obvious shorter notations, 
\barr
d\sigma^{A^\uparrow B \to C X} - d\sigma^{A^\downarrow B \to C X}
& = & \sum_{a,b,c,d} \> \int {d^2\bfk_\perp dx_a \, dx_b \over \pi z}
\label{diff} \\  
& & \mbox{\hskip-1.5truecm} \times \,
\left[\tilde f_{a/A^\uparrow}(x_a,\bfk_\perp)-
\tilde f_{a/A^\uparrow}(x_a,-\bfk_\perp)\right]\,f_{b/B}(x_b) \>
d\hat\sigma(\bfk_\perp)\>
D_{C/c}(z)\, .
\nonumber
\earr

The above new quantity (\ref{isiv2}), divided by twice the unpolarized 
$\bfk_\perp$ dependent distribution function $2\tilde f_{q/A}(x_a, \bfk_\perp)
= \tilde f_{q/A^\uparrow}(x_a, \bfk_\perp) + 
\tilde f_{q/A^\downarrow}(x_a, \bfk_\perp)$, can be regarded as a single 
spin asymmetry or analysing power for the $A^\uparrow \to a + X$ process;
similarly to the quark fragmentation analysing power,
Eq. (\ref{acq}), it may be different from zero for hadron
spin orientations perpendicular to the $A$-$q$ plane and
taking into account initial state interactions between
the two colliding hadrons. This quantity plays, for single spin asymmetries, 
the same role plaid by distribution functions in unpolarized processes. 
If we define the polarized number densities in terms of 
{\it distribution amplitudes} as 
\beq
\tilde f_{a, \la^{\,}_a/A^\uparrow}(x_a, \bfk_{\perp a}) =
\sumint_{X_A,\la_{X_A}}
|{\cal G}^{a/A}_{\la_{X_A}, \la^{\,}_a; \uparrow}(x_a,\bfk_{\perp a})|^2\, ,
\label{g}
\eeq
then we have, in the helicity basis,
\barr
\Delta^{\!N}\tilde f_{a/A^\uparrow}(x_a,\bfk_{\perp a}) &=&
\sumint_{X_A,\la_{X_A}} \sum_{\la^{\,}_a} \,2\,\mbox{Im} \, \left[
{\cal G}^{a/A}_{\la_{X_A}, \la^{\,}_a;+}(x_a,\bfk_{\perp a}) \,\,
{\cal G}^{a/A\,*}_{\la_{X_A},\la^{\,}_a;-}(x_a,\bfk_{\perp a}) \right]
\nonumber \\
&\equiv& 2 \, I^{a/A}_{+-}(x_a,\bfk_{\perp a}) \, ,
\label{iap}
\earr
which shows the non diagonal nature, in the helicity indices, 
of $I^{a/A}_{+-}(x_a,\bfk_{\perp a})$. 

Notice that, due to the odd $\bfk_\perp$ dependence of $I^{a/A}_{+-}$,
one has
\barr
& & \int d^2\bfk_\perp\,I^{a/A}_{+-}(\bfk_\perp)\,\frac{d\hat\sigma}
{d\hat t}(\bfk_\perp) \nonumber \\
& = & \int\!\!\!\!\!\!\!\!\!\!\!
\raisebox{-.6truecm}{$\scriptstyle (\bfk_{\perp})_x \,>\, 0$}
\!\!\!\!\!\!\!\! d^2\bfk_\perp\,I^{a/A}_{+-}(\bfk_\perp)\,
\left[\frac{d\hat\sigma}{d\hat t}(\bfk_\perp)-
\frac{d\hat\sigma}{d\hat t}(-\bfk_\perp)\right]\, ,
\label{kx}
\earr
where $d\hat\sigma/d\hat t(\bfk_\perp) - d\hat\sigma/d\hat t(-\bfk_\perp)
= {\cal O}(k_\perp / p_T)$, which clearly shows that this is a higher twist
effect.

A simple phenomenological model was developed in Ref.~9 
for the $\pup p \to \pi X$ process by assuming
that the dependence of $I^{a/p}_{+-}$ on $\bfk_{\perp a}$ 
is sharply peaked around an average value $k_{\perp a}^0$:
\beq
I^{a/p}_{+-}(x_a,k_x) = \frac{k_x}{M}\,\delta\left(|k_x|-
k_{\perp a}^0\right)N_a \,x_a^{\alpha_a}(1-x_a)^{\beta_a}
\label{ik0}
\eeq
with, from a fit to the data \cite{rob},
\beq
\frac{k_{\perp a}^0}{M} = 0.47 \> x^{0.68}\,(1-x)^{0.48} \,,
\label{ix}
\eeq
where $M$ is a hadronic mass scale, $M \simeq 1$ GeV/$c$. 

Inserting Eqs. (\ref{diff}), (\ref{kx}) and (\ref{ik0})
into Eq. (\ref{asym}) yields  
\beq
A_N = \frac{1}{M}\,
{\displaystyle \frac{ \sum N_a\,
\int dx_a dx_b \, k_{\perp a}^0 \, x_a^{\alpha_a}(1-x_a)^{\beta_a} \,
f_{b/p}(x_b) \, [d\hat\sigma(k_{\perp a}^0)-
d\hat\sigma(-k_{\perp a}^0)] \, D_{\pi/c}(z)/z}
{\sum \int dx_a dx_b ~ f_{a/p}(x_a) \, f_{b/p}(x_b)
d\hat\sigma \, D_{\pi/c}(z)/z}} \, .
\label{asyfin}
\eeq

In order to give numerical estimates of the asymmetry (\ref{asyfin}) 
we have taken $f_{q,\bar q,g/p}$ from Ref.~11, $D_{\pi/q,\bar q}$ 
from Ref.~12 and $D_{\pi/g}$ from Ref.~13. Given the very 
limited $p_T$ range of the data we have neglected the QCD $Q^2$ dependence 
of the distribution and fragmentation functions.
We have only considered contributions from $u$ and $d$ quarks inside the 
polarized proton, which certainly dominate at large $x_F$ values, that is 
$a = u,d$ in the numerator of Eq. (\ref{asyfin}). Instead, we have considered 
all possible constituents in the unpolarized protons, with $k_{\perp} =0$, 
and all possible constituent fragmentation functions. 

By using Eq. (\ref{ix}) into Eq. (\ref{asyfin}), 
together with the unpolarized $f_{a,b/p}$ and $D_{\pi/c}$ functions, we remain 
with an expression of $A_N$ still dependent on a set of 6 free parameters,
namely $N_a, \alpha_a$ and $\beta_a$ ($a = u,d$), defined in Eq. (\ref{ik0}). 
We have obtained a best fit to the data \cite{ada2}, shown in Fig. 1, 
with the following values of the parameters:
\beq
\begin{array}{lccc}
\rule[-0.6cm]{0cm}{1.3cm}
 \; & N_a & \alpha_a & \beta_a \\
\rule[-0.6cm]{0cm}{0.5cm}
 u \;\;\;   & 5.19  & 2.10       &  3.67     \\
\rule[-0.6cm]{0cm}{0.5cm}
 d \;\;\;   & \!\!\!\!\! -2.29 & 1.43       &  4.22     \\
\label{param}
\end{array}
\eeq
As the experimental data \cite{ada2} cover a $p_T$ range between 0.7 and
2.0 GeV/$c$ we have computed $A_N$ at a fixed value $p_T = 1.5$ GeV/$c$.

\begin{figure}
\centerline{
\epsfig{figure=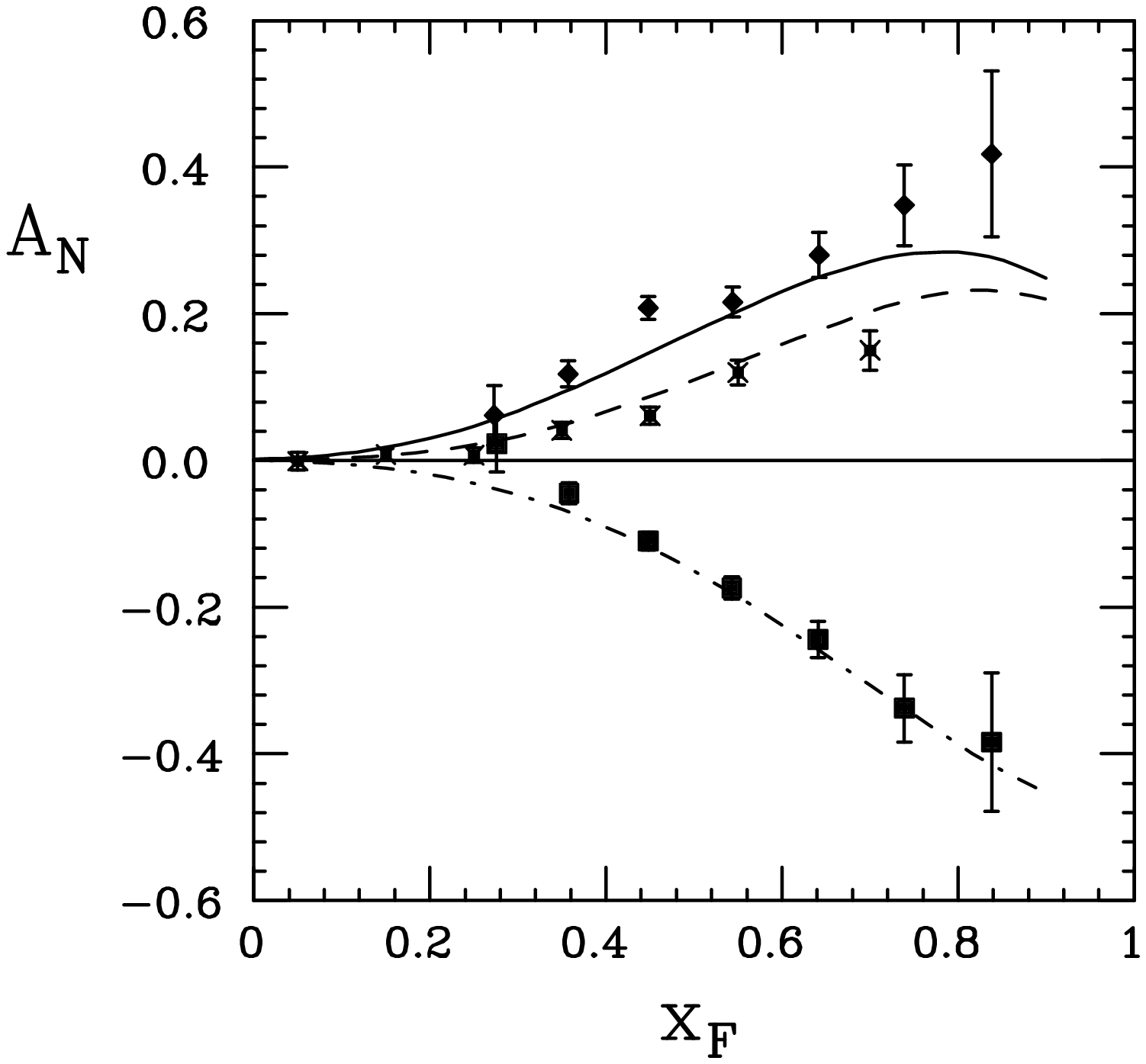,bbllx=160pt,bblly=272pt,bburx=492pt,%
bbury=604pt,width=7cm,height=7cm}}
\vskip1.5truecm
\fcaption{\begin{minipage}[t]{4.7in}
Fit of the data on $A_N$ \cite{ada2}, with the parameters
given in Eq. (\ref{param}); the upper, middle and lower sets of data 
and curves refer respectively to $\pi^+, \pi^0$ and $\pi^-$.
\end{minipage}}
\label{fig}
\end{figure}

Notice that the above values (\ref{param}) are very reasonable indeed; 
actually, apart from an overall normalization constant, they might even have 
been approximately guessed. The exponents $\alpha_{u,d}$ and $\beta_{u,d}$ are 
not far from the very na\"{\i}ve values one can obtain by assuming, as somehow  
suggested by Eqs. (\ref{g}) and (\ref{iap}), that  $I^{a/p}_{+-}(x_a) 
\sim \sqrt {f_{a,+/p,+}(x_a)f_{a,-/p,+}(x_a)}$, where 
$f_{a,+(-)/p,+}(x_a)$ denotes, as usual, the 
number density of quarks with the same (opposite) helicity as the parent
proton. Also the relative sign and strength of the normalization constants
$N_u$ and $N_d$ turn out not to be surprising if one assumes that there 
might be a correlation between the number of quarks at a fixed value of 
$\bfk_{\perp a}$ and their polarization: remember that, according to $SU(6)$, 
inside a proton polarized along the $\hat{\bfy}$ direction, $P_y=1$, one has
for valence quarks $P_y^u = 2/3$ and $P_y^d = -1/3$.

It might appear surprising to have approximately opposite values for the 
$\pi^+$ and $\pi^-$ asymmetries, as the data indicate, and a large  
positive value for the $\pi^0$; one might rather expect $A_N \simeq 0$ for a 
$\pi^0$. However, this can easily be understood from Eq. (\ref{asyfin}) which 
we simply rewrite, for a pion $\pi^i$, as  $A_N^i = {\cal N}^i/{\cal D}^i$, 
if one remembers that from isospin symmetry one has:
\beq
D_{\pi^0/c} = \frac{1}{2}\left( D_{\pi^+/c} + D_{\pi^-/c} \right)\,.
\label{iso}
\eeq

Eq. (\ref{asyfin}) shows that the relation 
(\ref{iso}) also holds for ${\cal N}^0$ and ${\cal D}^0$, so that
\beq
A^0_N = \frac{{\cal N}^+ + {\cal N}^-}{{\cal D}^+ + {\cal D}^-} =
A^+_N \;\, \frac{ 1 + \frac{\displaystyle A^-_N}{\displaystyle A^+_N}
\frac{\displaystyle {\cal D}^-}{\displaystyle {\cal D}^+}}
{1 + \frac{\displaystyle {\cal D}^-}{\displaystyle {\cal D}^+}} \,\cdot
\label{a0}
\eeq
It is then clear that $A^-_N \simeq - A^+_N$ implies $A^0_N \simeq 0$ only 
if $\cal D^- \simeq \cal D^+$, {\it i.e.} if the unpolarized cross-sections 
for the production of a $\pi^-$ and a $\pi^+$ are approximately equal.
This is true only at $x_F \simeq 0$. At large $x_F$ the minimum value of 
$x_a$ kinematically allowed increases and the dominant contribution to 
the production of $\pi^+$ and $\pi^-$ comes respectively from $f_{u/p}(x_a)$ 
and $f_{d/p}(x_a)$ [see the denominator of Eq. (\ref{asyfin})]. It is 
known that $f_{d/p}(x_a)/f_{u/p}(x_a) \to 0$ when $x_a \to 1$; this implies
that ${\cal D}^-/{\cal D}^+$ decreases with increasing $x_F$, so that {\it at 
large} $x_F$ we have ${\cal D}^-/{\cal D}^+ \ll 1$ and  $A^0_N \simeq 
A^+_N( 1 - 2{\cal D}^-/{\cal D}^+) \simeq  A^+_N$. Such a trend emerges both 
from the experimental data and our computations.

We have discussed how to compute single spin asymmetries in large $p_T$ 
inclusive production within the framework of the factorization theorem and 
perturbative QCD, showing that single spin effects of order $k_\perp/p_T$ 
can be related to non perturbative intrinsic properties of quark 
fragmentations and distributions, {\it i.e.} the {\it quark fragmentation 
analysing power} and the {\it quark distribution analysing power}; the 
former was first suggested by Collins \cite{col2}
and the latter by Sivers \cite{siv1,siv2}. 
We have shown a possible description of the single spin asymmetries observed 
in $p^\uparrow p \to \pi X$ via the intrinsic $\bfk_\perp$ effects in the 
quark distribution functions.

A definite test and a better evaluation of these non perturbative 
properties requires further and more refined applications of the same idea 
and more theoretical work. In particular one might consider the following 
processes:

\begin{description}
\item{--}
$\bar p^\uparrow p \to \pi X$; a straightforward application of the model 
previously described, with the same set of parameters, gives results  
comparable with some existing data \cite{ada1}.
\item{--}
$p^\uparrow p \to \gamma X$; in such a case there cannot be any fragmentation
effect and one should be able to single out the importance of the Sivers
effect and the quark distribution analysing power.
\item{--} 
$\ell p^\uparrow \to h X$; in such a case, the inclusive production of 
a hadron in DIS, there should be no effect from the quark distribution
analysing power, as any initial state interaction would be of higher 
electromagnetic order and negligible. It should then allow an evaluation
of the Collins effect.
\item{--}
$p^\uparrow p \to \pi X$, with a correct inclusion of both the Collins and
Sivers effect.
\end{description}

Let us finally mention that we have assumed the validity of the QCD 
factorization theorem also at higher twist. This has been discussed in
the literature: an approach similar to that discussed here \cite{noiplb} 
has been advocated, in the operator language, by Qiu and Sterman \cite{qiu} 
who use generalized factorization theorems valid at higher twist
and relate non zero single spin asymmetries in $p\,p$ collisions to the 
expectation value of a higher twist operator, a twist-3 parton distribution,
which explicitly involves correlations between the two protons and combines 
quark fields with a gluonic field strength. However, they still consider
only collinear partonic configurations so that, in order to obtain non 
zero results, they have to take into account the contributions of higher
order elementary interactions. Some more theoretical work is still necessary.

%

\goodbreak
\vskip20pt
\nd 
\end{document}